\documentclass[twocolumn,showpacs,preprintnumbers,amsmath,amssymb]{revtex4}
\usepackage{graphicx}
\usepackage{dcolumn}
\begin{document}


\title
{Density pattern in supercritical flow of liquid $^4$He}

\author{F.~Ancilotto$^{1}$, F.~Dalfovo$^{2,3}$, L.P.~Pitaevskii$^{2,4}$ and
F.~Toigo$^{1}$}
\affiliation{
$^{1}$ INFM and Dipartimento di Fisica, Universit\`{a}
di Padova, Via Marzolo 8, 35131 Padova, Italy \\
$^{2}$ BEC-INFM, Istituto Nazionale per la Fisica della Materia, Trento, Italy\\
$^{3}$ Dipartimento di Matematica e Fisica, Universit\`{a} Cattolica, 
Via Musei 41, 25121 Brescia, Italy\\
$^{4}$ Dipartimento di Fisica, Universit\`{a} di Trento, 38050 Povo, Italy \\
and Kapitza Institute for Physical Problems, ul. Kosygina 2, 
117334 Moscow, Russia }
 
\date{\today}

\begin{abstract}
A density functional theory is used to investigate the instability 
arising in superfluid $^4$He as it flows at velocity $u$ just above the 
Landau critical velocity of rotons $v_c$. Confirming an early theoretical
prediction by one of us [JETP Lett. {\bf 39}, 511 (1984)], we find that 
a stationary periodic modulation of the density occurs, with amplitude 
proportional to $(u-v_c)^{1/2}$ and wave vector equal to the roton wave 
vector. This density pattern is studied for supercritical flow both in 
bulk helium and in a channel of nanometer cross-section.
\end{abstract}
\maketitle

According to Landau criterion superfluid motion of $^4$He moving 
with velocity $u$ is possible only if $u<v_c$, where the critical 
velocity $v_c$ is equal to the slope of the tangent to the roton 
part of the spectrum ($v_c\sim 56$ m/s). As the Landau critical 
velocity $v_c$ is reached, the liquid becomes unstable against 
a spontaneous excitation of rotons. Reaching the roton critical 
velocity is difficult in practice since other type of excitations, 
e.g. quantized vortices, are produced in bulk $^4$He well below $v_c$. 
However, the occurrence of vorticity could be suppressed by allowing 
$^4$He to flow through very narrow channels. In fact, the critical 
velocity for the creation of vortex pairs in a channel of diameter 
$D$ is $v_c^{vortex} \sim (\hbar/M D) \ln(D/\xi )$ \cite{feynman}, 
where $\xi $ is the $^4$He healing length ($\xi \sim 1$ \AA), so that 
it can exceed the roton critical velocity for channels of nanometer 
size \cite{note1}. 

Several years ago a theoretical prediction was made by one of us 
\cite{pita} that superfluid $^4$He flowing at a velocity greater than 
the Landau critical velocity $v_c$ should undergo a phase transition 
from a spatially homogeneous state to a layered state characterized 
by a periodic density modulation in the direction of motion. Such 
a modulation is stationary in the frame moving with the fluid and 
has a characteristic wavelength $\lambda = 2\pi \hbar/p_c\sim 3.58$~\AA, 
where $p_c$ is the roton momentum. This prediction was derived within 
a simplified model describing a weakly interacting roton gas with 
coupling constant $g$. The nature of the transition was found to 
depend on the sign of $g$: if $g>0$ ($g<0$) the transition is predicted 
to be continuous (discontinuous). In Ref.~\cite{pita} 
the estimate $g=2\times 10^{-38}$ erg cm$^3$ \cite{exptg} was used 
and the amplitude of the density modulations was found to be 
\cite{oldmisprint}
\begin{equation}
\frac{\Delta \rho}{\rho _0}=
2 \left( \frac{|A|^2(u-v_c)p_c}{\rho _0\,g} \right)^{1/2} 
\label{eq:deltarho}
\end{equation}
where $\rho_0$ is the bulk density and $|A|^2 \delta(\hbar \omega -
\epsilon(p_c))$ is the roton contribution to the dynamic structure 
factor $S(q,\omega)$. In Ref.~\cite{pita} the latter was estimated 
by ignoring the multiphonon part of $S(q,\omega)$ and using the
$f$-sum rule. A better estimate can be extracted from neutron 
scattering experiments \cite{cowley}, where one finds $|A|^2 \simeq
0.9$. Inserting this value in Eq.~(\ref{eq:deltarho}) one gets
\begin{equation}
\Delta \rho/\rho _0 \simeq 3 [(u-v_c)/v_c]^{1/2} \; .
\label{eq:deltarho2} 
\end{equation}

The occurrence of this stationary nonuniform state originates from 
the presence of a pronounced minimum at $p=p_c$ in the bulk $^4$He 
spectrum, $\epsilon(p)$, and is similar to the structural phase 
transition in crystals induced by the softening of phonon 
frequencies with some defined wavelength. It is also worth
mentioning that similar periodic modulations in the $^4$He density 
profile near a moving impurity have been observed in recent 
computer simulations \cite{berloff}, although not explained in 
terms of this instability. 

In this work we investigate the occurrence of density patterns in 
the supercritical $^4$He flow by performing density functional (DF) 
calculations. We consider a uniform flow in bulk liquid (with no 
vorticity) as well as in a nanochannel. We use the DF approach 
proposed in Ref.~\cite{dupont} and later improved in Ref.~\cite{prica}, 
which gives a quite accurate description of inhomogeneous 
configurations of liquid $^4$He at $T=0$.
The energy of the system is expressed as:
\begin{equation}\label{StatFunc}
E_0[\rho ]=\frac{1}{2 M}
\int d{\bf r} \left( \nabla
\sqrt{\rho}\right)^2
+\int d{\bf r} \,E_c({\bf r}) \; . 
\end{equation}
The explicit form of the correlation energy $E_c$ is given in 
Ref.~\cite{prica}. The {\it static} equilibrium profile 
$\rho ({\bf r})$ in an arbitrary external potential can be obtained 
by minimizing the functional $E_0[\rho ]$ with respect to density 
variations, subject to the constraint of a constant number of 
atoms. The {\it dynamics} can be studied as well by means of the 
time dependent DF method, with the DF proposed in 
Ref.~\cite{prica} playing the role of the effective Hamiltonian 
driving the time evolution of the system. In the dynamical case, the 
functional contains an  explicit dependence upon the local current 
density field ${\bf j({\bf r})}$ through a phenomenological term 
which accounts not only for the usual hydrodynamic current density 
but also for non-local ``backflow" effects. The resulting DF (named 
Orsay-Trento Functional), which will be used in our calculations,
has the following form: 
\begin{equation}
E[\rho,{\bf v}]
=E_0[\rho ] + \int d{\bf r} H_J \; . 
\end{equation}
An appealing feature of the above functional, which turns out to 
be essential to perform accurate time dependent DF calculations
\cite{prica,giacomazzi}, is that it reproduces quantitatively not only a number 
of static properties, but also the observed phonon-roton spectrum 
of bulk $^4$He.

The minimization of the above density-current functional, subject 
to the constraint of a fixed number of $^4$He atoms and of fixed 
total momentum, can be done in practice by evolving in the imaginary 
time domain a non-linear Schr{\"o}dinger equation for the order parameter 
$\Psi ({\bf r})$, where the Hamiltonian operator is given by 
$H=- 1/(2M) \nabla^2 + U[\rho,{\bf v}]$. The effective potential 
$U$ is defined in terms of the variational derivative of the energy 
functional, and its explicit expression can be found in 
Ref.~\cite{giacomazzi}. From the knowledge of the complex 
wavefunction $\Psi\equiv \phi e^{i\Theta }$
one can get immediately the density $\rho({\bf r})=\phi^2$ and 
the velocity field ${\bf v}({\bf r})= (1/M) \nabla \Theta$.
Since we are interested in stationary states of $^4$He in the 
presence of a uniform flow, we minimize the above functional in 
the frame of reference moving with the liquid, which we assume 
to flow with some given velocity $u$ along the $x$-axis:
The Hamiltonian density $H$ thus acquires an additional term
$H^\prime =H-u \hat{P}_x$, $\hat{P_x}$ being the $^4$He total 
momentum component along the direction of motion.

\begin{figure}[tbh!]
\centerline{\includegraphics*[clip,width=7.5cm,angle=0]{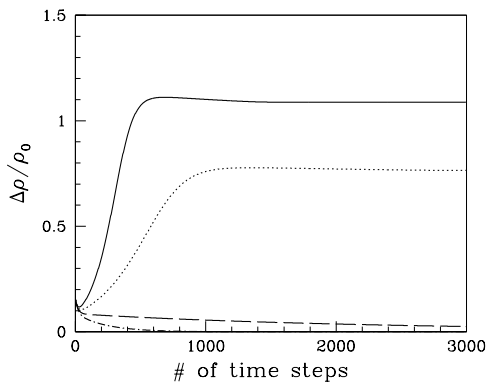}}
\caption{Amplitude of the density modulation along the direction of
$^4$He motion, computed during the functional minimization.
Solid line: $u=1.29\,v_c$, dotted line: $u=1.14\,v_c$,
dashed line: $u=0.99\,v_c$, dash-dot line: $u=0.84\,v_c$}
\label{Fig1}
\end{figure}

First, we address the problem of the Landau roton instability
in bulk. As discussed above, we expect that when $u>v_c$ the 
uniform density configuration is not stable, but it is instead a 
metastable state corresponding to a saddle point of the energy 
landscape of $^4$He. In our case, the systems is allowed to reach 
the lowest energy configuration by following the (dissipative)
imaginary-time evolution.  The
calculation is performed in a periodically repeated supercell where 
the size of the cell along the $x$-direction (which we take as the 
direction of $^4$He motion) is $L$. Our procedure to trigger the 
instability is the following: we start with the uniform system in 
the moving frame of reference and slightly perturb the (uniform) 
density with a sinusoidal modulation with a small arbitrary 
amplitude and with a wavelength $\lambda $ allowed by the periodic 
boundary conditions in $L$. We then minimize the functional in 
the  frame of reference moving with some chosen velocity $u$,  
with the only constraint of a constant number of $^4$He atoms.
If $L$ or $\lambda$ are not a multiple of the characteristic 
wavelength $\lambda_c \equiv (2\pi)/k_c$ ($k_c\equiv p_c/\hbar$ 
being the Landau critical wave-vector) then, irrespective 
of the initial perturbation and of the particular value chosen 
for $u$, the perturbing modulation rapidly smoothes out during 
the minimization, and the uniform liquid state is recovered as the 
minimum energy configuration.

\begin{figure}[tbh!]
\centerline{\includegraphics*[clip,width=7.5cm,angle=0]{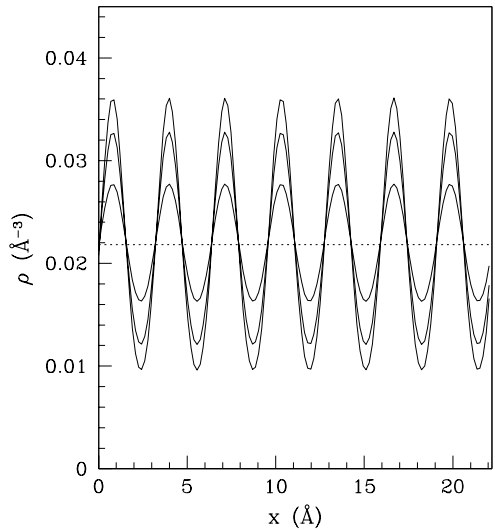}}
\caption{Density profiles along the direction of $^4$He flow $x$-axis.
The three profiles have been calculated, in order of increasing
amplitude, with $u=1.07\,,1.22 \,,1.37\,v_c$, respectively.}
\label{Fig2}
\end{figure}

The Landau instability shows up when the size of the cell is such 
to accomodate an integer number of characteristic wavelengths ($L=7 
\lambda_c$). In this case we indeed find that there exists a threshold 
velocity $v_c$ separating two regimes. If $u>v_c$, the stationary 
state is characterized by a density with a periodic modulation of 
wavelength $\lambda _c$ and with an amplitude  depending on $u$. 
On the contrary, when  $u<v_c$ the initial modulation is rapidly 
smeared out during the minimization, and again one finds that the 
density of the stationary state is uniform. The critical velocity 
is found to be $v_c\sim 58$ m/s, which coincides with the minimum 
value of $\epsilon(p)/p$ predicted by the same DF and is also very close
to the value of the Landau critical velocity of rotons
as obtained from the experimental phonon-roton spectrum.
This behavior is summarized in Fig.~1, where we plot the evolution 
of the amplitude of the density modulation as it varies during the 
minimization procedure, for four different $^4$He velocities: the 
two upper lines have $u>v_c$, whereas the two lower lines have $u<v_c$.
Note the critical slowing down for values of the $^4$He velocity
close to the critical value $v_c$, where a very long imaginary-time 
evolution is required to converge towards the equilibrium stationary 
state. 

Different stationary density profiles along the direction of $^4$He 
motion, corresponding to different values of $u>v_c$, are shown in Fig.~2. 
The average value of each curve corresponds to the saturation density 
of bulk $^4$He, $\rho_0=0.0218$ \AA$^{-3}$. A fit to the calculated
points shows that their shapes, at least for values of $u$ not too 
large, is almost exactly sinusoidal, i.e, $\rho (x)= \rho _0 [ 1+ (\Delta 
\rho / \rho _0) \sin(k_cx)]$.  In Fig.~3 we also show the $x$-component 
of the calculated $^4$He velocity  ${\bf v}({\bf r})=(1/M) \nabla \Theta$,
in units of $v_c$, for the same states of Fig.~2. Note the oscillating 
character of the velocity, in phase with the density modulation, and 
the large amplitude of oscillations, which becomes more asymmetric as 
the velocity increases. The spatial average of the velocity profiles 
shown in Fig.~3 is zero, as expected.

\begin{figure}[tbh!]
\centerline{\includegraphics*[clip,width=7cm,angle=0]{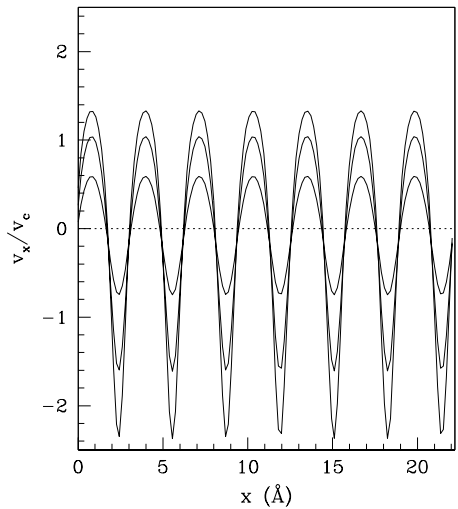}}
\caption{Velocity profile along the direction of $^4$He flow $x$-axis.
Same values of $u$ as in Fig.~2}
\label{Fig3}
\end{figure}

The main result of this work is summarized in Fig.~4 where we show 
the behavior of the amplitude $\Delta \rho/\rho _0$ for $u>v_c$. We 
find that the law $\Delta \rho / \rho _0 = 1.01 [ (u-v_c)/ v_c]^{1/2}$ 
(solid line) very nicely fits the numerical results (points). The 
velocity dependence is thus the same as in Eq.(\ref{eq:deltarho2})
except for the different numerical coefficient. Our DF calculations 
are consistent with a repulsive (positive $g$) roton-roton interaction.
Using Eq.~(\ref{eq:deltarho}) and the fitting coefficient $1.01$, we 
find $g \simeq 1.8 \times 10^{-37}$ erg cm$^3$. It is worth stressing 
that direct measurements of $g$ are not available and previous 
theoretical estimates significantly differ both in magnitude and 
sign (see, for instance, \cite{nagai,pistolesi} and references 
therein). 
 
\begin{figure}[tbh!]
\centerline{\includegraphics*[clip,width=7cm,angle=0]{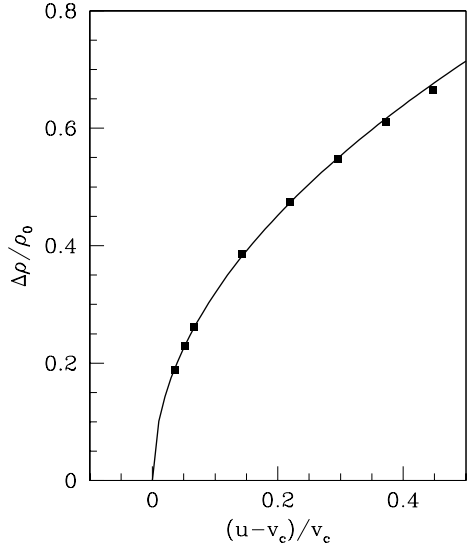}}
\caption{Amplitude of the density modulation a function of the 
fluid velocity. Points: DF results. Line: fitting function 
$1.01 [(u-v_c)/v_c]^{1/2}$}
\label{Fig4}
\end{figure}

Now we investigate the motion of liquid $^4$He in a narrow channel 
of nanometer transverse dimensions. We consider liquid $^4$He 
confined between two infinitely extended, weakly attractive planar 
surfaces separated by a very small distance, $\sim \,50$ \AA. We 
model the two surfaces with an external potential which mimics the 
adsorption properties of the Rb surface, which is the weakest surface 
which is wet at $T=0$ by liquid $^4$He \cite{anci}. The number of 
$^4$He atoms in the system is chosen in such a way that, when the 
$^4$He is at rest, the equilibrium density near the center of the 
channel reaches the value corresponding to the saturation density 
of bulk $^4$He, $\rho_0$.

\begin{figure}[tbh!]
\centerline{\includegraphics*[clip,width=6cm,angle=0]{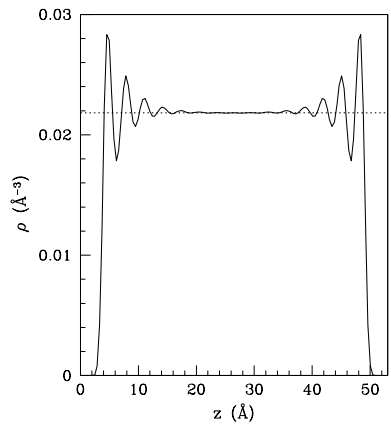}}
\caption{Density profile across the channel section}
\label{Fig5}
\end{figure}

In Fig.~5 we show the density profile along the $z$-direction, i.e. 
across the channel, for $u=0$. The $^4$He density decreases rapidly 
to zero near the solid surfaces on both sides of the channel due to the 
$^4$He-Rb interaction. The same interaction is also responsible for the
density oscillations near the walls. The dotted  line shows the value 
of the bulk saturation density $\rho _0$. Fig.~6 shows a contour plot 
of the density in the $xz$-plane for the stationary state developed at 
$u=1.22\,v_c$. The complex pattern near the walls is again due to 
the $^4$He-Rb interaction. However, the dominant feature is the density 
modulation along $x$ in the central part of the channel. This sinusoidal 
oscillation coincides, for the same value of $u$, with the one that
we already obtained in bulk $^4$He.

\begin{figure}[tbh!]
\centerline{\includegraphics*[clip,width=8cm,angle=0]{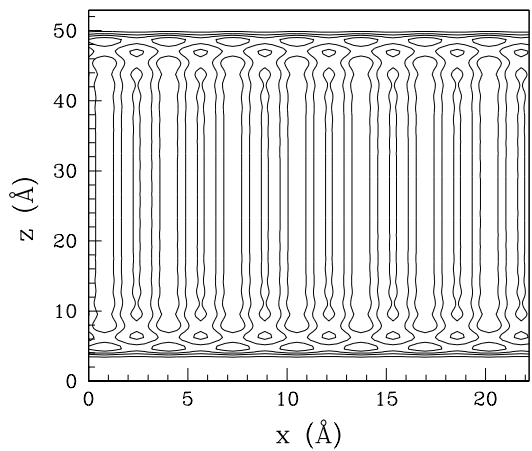}}
\caption{Contour plots of the density along the channel}
\label{Fig6}
\end{figure}

Our DF calculations support the predictions of Ref.~\cite{pita}
on the occurrence of a density pattern in the supercritical 
flow just above $v_c$ and in the absence of vorticity. Due to the
short wavelength of the density modulations, of the order of the 
atomic spacing, its direct observation, with X-rays for instance, 
might be difficult. Indirect evidences of the density modulations
could however be measurable, for example throught their possible 
effects on transport properties. Recently, He adsorption
within a regular porous medium called FSM-16, has been 
studied \cite{wada}. This silica-based material is characterized 
by ordered arrays of long, uniform pores, with diameters ranging 
from $1.5$ to $10$ nm. When $^4$He is adsorbed within the pores, 
1-2 solid-like layers are expected to form, coating the internal 
walls of the pores, leaving however room for additional $^4$He 
in the liquid state. A pressure gradient between two open ends of
an array of pores could in principle be used to force liquid
$^4$He to move through this system, until it is expelled from 
the pore end. If during this process the critical velocity is 
reached, then the occurrence of the above described density pattern 
might induce the fragmentation of the ejected liquid filament 
into regularly distributed nanodroplets. A similar process might
occur in the experiments of Ref.~\cite{toennies}, where 
liquid $^4$He is discharged into vacuum through a micrometer nozzle.
The structure of the ejected filament was interpreted in terms
of a Rayleigh instability, but the occurrence of density patterns 
near the nozzle could also play a role \cite{rossi}. In this 
perspective, the effects of density modulations in $^4$He 
supercritical flow in this type of experiments deserve further
investigations. Finally, it is worth mentioning that a similar 
phenomenon may occur in Bose-Einstein condensed gases with 
dipole-dipole interactions \cite{giovanazzi}.

\acknowledgments
We thank Maurizio Rossi for useful conversations.


\begin{thebibliography}{12}
\bibitem{oldmisprint} We note on passing that the factor $2$ was 
erroneously missing in ref.\protect\cite{pita}.
\end{thebibliography}
\end{document}